\documentclass[a4paper]{article}
\usepackage[margin=0.5in]{geometry}
\usepackage{amsmath,amsfonts,amssymb,bbm}
\usepackage{graphicx,graphics}
\DeclareGraphicsExtensions{.pdf,.eps}
\usepackage{xspace}
\usepackage[latin1]{inputenc}
\usepackage[
	backend=bibtex,
	sorting=none,
	firstinits=true,
	maxbibnames=99,minbibnames=99,
	style=numeric-comp
	]{biblatex}
\bibliography{mine}
 \newtheorem{thm}{Theorem}
 \newcommand{\proof}{\noindent {\bf Proof: }}
\makeatletter 
\newcommand{\ket}     [1] {\ensuremath{\left|#1\right\rangle}}

\newcommand{\isotope} [2] {\ensuremath{{}^{#2}\!{\mathrm{#1}}}} 
\newcommand{\multisotope} [3] {\ensuremath{{}^{#2}\!{\mathrm{#1}_{#3}}}}
\newcommand{\Alp}{\ensuremath{\mathrm{\alpha}}}

\newcommand{\XRA}{\xrightarrow}

\newcommand{\Cthree}{\ensuremath{\mathrm{C3}}}
\newcommand{\Cfour}{\ensuremath{\mathrm{C4}}}
\newcommand{\Cfive}{\ensuremath{\mathrm{C5}}}

\newcommand{\Htwo}{\ensuremath{\mathrm{H2}}} 
\newcommand{\Htwoa}{\ensuremath{\mathrm{H2\alpha}}} 
 
\newcommand{\Hfour}{\ensuremath{\mathrm{H4}}}

\def\eps  {\ensuremath{\varepsilon}}
\def\sec  {\ensuremath{\mathrm{s}}}
\def\msec {\ensuremath{\mathrm{ms}}}
\def\Hz   {\ensuremath{\mathrm{Hz}}}
\def\kHz  {\ensuremath{\mathrm{kHz}}}

\def\FF   {\mathcal{F}}          
\def\TT   {\mathcal{T}}

\def\BB   {\ensuremath{\mathcal{B}}}
\def\RR   {\ensuremath{\mathcal{R}}\xspace}
\def\QQ   {\ensuremath{\mathcal{Q}}\xspace}
\def\DD   {\ensuremath{\mathcal{D}}\xspace}

\def\qed   {\mbox{ }~\hfill~$\Box$}

\def\Cone  {\ensuremath{\mathrm{C1}}} 
\def\Ctwo  {\ensuremath{\mathrm{C2}}} 
\def\Tone  {\ensuremath{\mathrm{T_1}}} 
\def\Tonec {\ensuremath{\mathrm{T_1 (comp) }}} 
\def\Toner {\ensuremath{\mathrm{T_1 (reset) }}}

\def\Trun  {\ensuremath{\mathrm{T_{run}}}} 
\def\Twait {\ensuremath{\mathrm{T_{WAIT}}}\xspace} 
\def\ppm   {\ensuremath{\mathrm{ppm}}}

\newcommand\super [1]   {\ensuremath{^{\mathrm{#1}}}\ }

\def\nd   {\super{nd}}

\def\th   {\super{th}}

\begin{document}
\title{Prospects and Limitations of Algorithmic Cooling}
\author{Gilles Brassard$^1$, Yuval Elias$^2$, Tal Mor$^2$, and Yossi Weinstein$^2$ 
\\~\\
$^1$~D{\'e}partement IRO, Universit{\'e} de Montr{\'e}al, Montr{\'e}al, H3C 3J7~~Canada 
\\
$^2$~Department of Computer Science, Technion 32000, Haifa, Israel
}
\maketitle
\abstract{
Heat-bath algorithmic cooling (AC) of spins 
is a theoretically powerful 
effective cooling approach, 
that (ideally) cools spins with low polarization  exponentially 
better than cooling by reversible entropy manipulations alone.
Here, we investigate the limitations and prospects of AC\@. 
For non-ideal and semioptimal AC, we study the impact of finite
 relaxation times of reset and computation spins on the achievable effective cooling. We 
derive, via simulations, 
the attainable cooling levels for given ratios of relaxation times using two semioptimal practicable algorithms. 
We expect this analysis to be valuable for the planning of 
future experiments. 
For ideal and optimal AC, we make use of lower 
bounds on the number of required reset steps, based on entropy considerations,
to present important consequences of using AC as a tool for improving
signal-to-noise ratio in liquid-state magnetic resonance spectroscopy.
We discuss the potential use of AC for noninvasive clinical diagnosis and drug monitoring, where it may have significantly lower specific absorption rate (SAR) with respect to currently used methods.
} 
\section{Introduction}\label{sec:intro}
%
%
Algorithmic cooling (AC)~\cite{BMRVV02} is useful for initiating a spin-half
system, or more generally, a quantum bit (qubit) or quantum digit (qudit) 
system~\cite{FLMR04,SMW05,SMW07,CML11,Blank13,BCC+07}, 
and for potentially improving signal-to-noise ratio (SNR) 
in liquid-state magnetic resonance 
spectroscopy~\cite{FLMR04,POTENT,AAC-pat,EGMW11,AEMW-1,AEMW-2}. 
AC of spins allows 
effective cooling of a single or several spins far beyond 
close-system cooling (reversible entropy manipulations~\cite{Sorensen89, SV99}).
Such effective spin-cooling is used to reach an effective temperature, smaller
(or even much smaller) than the actual temperature of 
the environement, and
then the spins relax back
to the temperature of the environement, in a typical time scale
called the relaxation time $\Tone$. 

See~\cite{FLMR04} for a detailed presentation of the bound of $\sqrt{n}$ on
 cooling $n$ spins using  
reversible entropy manipulations and the exponential bypass achievable by AC\@.
Exponential cooling was shown for practicable AC (PAC) algorithms~\cite{FLMR04},
and improved exponential cooling was shown for optimal 
algorithms\footnote{The word \emph{optimal} here refers to the achieved spin temperature. The partner 
pairing algorithm is also most efficient with respect to the number of reset steps, see below. } --- the
partner-pairing algorithm~\cite{SMW05,SMW07} and
all-bonacci~\cite{EFMW06,EFMW07}. 
Note that the analysis here is relevant for high-temperature
(e.g. room temperature) liquid NMR, 
hence for a very small initial polarization
bias, $\eps \ll 1$ (typically
around $10^{-5}$). The exponential cooling mentioned above is 
an approximation which is true as long as the final bias
also satisfies that it is much smaller than~1.
In principle, AC is a microscopic-scale quantum heat 
pump~\cite{FLMR04,RMM07,HRM07,WHRSM08,BP08,LPS10,DRRV11,Renner12},
and it is found useful recently for various 
purposes, see~\cite{BPM+11,SBR+11,Lloyd14,XYX+14}, 
in addition to the abovementioned ones. 

AC is based on the presence, within the same spin system, of
two types of spins: spins that interact rapidly with the environment, termed
\emph{reset spins} and slow-relaxing \emph{computation spins}, which can store
and exploit the enhanced polarization. Ideally, the reset spins repolarize
infinitely faster than the computation spins. 
Let \RR denote the ratio between the relaxation time of the 
computation spins 
($\Tonec$),
and the relaxation time of the reset spins 
($\Toner$), namely,  
$\RR \triangleq \Tonec/\Toner$.
The ideal scenario where 
$\RR \rightarrow \infty$ 
is assumed in most theoretical papers, e.g.,  
in~\cite{BMRVV02, FLMR04, SMW05, EFMW06, EFMW07, SMW07,EMW11}. 
As suggested in~\cite{EMW11}, it is important to deal with a realistic scenario where \RR is finite, and as suggested in~\cite{EGMW11} it is interesting to 
compare AC to other methods used to improve the SNR in magnetic resonance 
spectroscopy. 

%
%
The search for better algorithms has lead to both optimal  
AC~\cite{SMW05, SMW07}, 
and practicable AC (PAC)~\cite{FLMR04, EFMW06, SMW07, EFMW07, EMW11}. 
The optimal 
``partner-pairing algorithm'' (PPA)~\cite{SMW05, SMW07} was shown to require
the least amount of reset steps
(namely, WAIT steps), however, its unitary steps are extremely
complicated (when written using 2-3 qubit gates).
In contrast, PAC~\cite{FLMR04} and semioptimal PAC (SOPAC)~\cite{EMW11}, 
which use very simple quantum gates in-between reset
steps,    
cool various spin-systems to 
a much lesser extent than the PPA, for any given number of reset steps. 
For any odd number of spins, $n=2j+1$, the SOPAC algorithm called mPAC
(see~\cite{EMW11})
cools, ideally, the most significant bit
 by a factor of $\left(2-\frac{1}{2^m}\right)^{j}$,
where $m$ denotes the number of compressions at each recursive level, and $j$ denotes the purification level. 
For $m=2,$ the exponential
 cooling of $1.75^j$ provided by 2PAC is much greater
 than the exponential cooling~\cite{FLMR04} provided 
by 1PAC\@.

Optimal cooling (via PPA)
could approach a cooling factor of $2^{n-2}$, and for any non-trivial cooling
 level (e.g., $n \gtrsim 6$) achievable by both the PPA and 2PAC, the PPA uses 
 much fewer reset steps. Another SOPAC algorithm, named mFib, also presented in~\cite{EGMW11}, belongs to a family of algorithms containing also $\delta$-Fib,
 Tribonacci and others~\cite{SMW07,EFMW06,EFMW07}, the bonacci AC, that deserve more attention and are reviewed and extended in section~\ref{sec:bonacci}.

%
%
When AC is moved from theory to
experiment~\cite{POTENT,BMR+05,RMBL08,EGMW11,Atia-MSc-Thesis,AEMW-1,AEMW-2}
the cooling, of course, is limited due to the realistic reset time.
The reset spins do not relax infinitely faster 
than the computation 
spins, and one cannot assume anymore that computation spins do not reheat
during reset steps. In real life one must account for a
finite ratio between the relaxation times of the computation and reset spins.
In sections~\ref{sec:account-for-T1} and~\ref{sec:AC-bound-resets} 
we analyze two algorithms, the PPA and 2PAC, 
that ideally provide exponential cooling,
in order to present two limitations of AC and understand their consequence.

%
%
In section~\ref{sec:account-for-T1} 
we present numerical simulation of the practicable algorithm 2PAC,
applied onto 7 spins, for $ 100 \le \RR \le 10,000$. 
We use a simplified model  of relaxation 
that prevents the need of simulating the relaxation of all correlation terms
 between spins, and is therefore suitable for a greater number of spins. 
We show precisely 
how the exponential advantage over reversible cooling is lost when the number
of spins is increased or \RR is decreased.

%
%
One may wonder whether a more efficient algorithm such as the PPA can yield 
much better results.
In section~\ref{sec:AC-bound-resets} we take into account that each reset step
takes a finite time (without discussing the value of $\RR$).
We employ entropy
 considerations (following~\cite{SMW07})
that provide lower bounds on the number of reset steps required
(ideally) for the PPA and hence also for any cooling algorithm. 
This result clarifies that there are 
limitations on AC for \emph{any}
cooling algorithm, which makes one wonder whether AC is useful for improving the SNR in magnetic resonance spectroscopy. 

%
%
In section~\ref{sec:prospects} we present two important cases in which AC might still be found very useful for that purpose in magnetic resonance spectroscopy for noninvasive clinical diagnosis. For instance, AC may supply the same quality of measurements as the currently used methods but with significantly lower amounts of RF radiation applied to the patient.

\section{Bonacci algorithms}\label{sec:bonacci}
It is interesting to consider the family of cooling algorithms which 
asymptotically achieve a Fibonacci-like sequence as the final
biases~\cite{SMW07,EFMW06}. 
The first example is the Fibonacci algorithm~\cite{SMW07,EFMW06,EMW11} 
which is recursively defined as \\
\emph{Fibonacci, $m$-Fibonacci and $\delta$-Fibonacci: Run $\FF(n,n)$}
\begin{equation}\label{eq:Fibonacci-definition}
\FF(n , k)=\left[\FF(n,k-1)\; 3BC(k)\right]^{m_{n,k}}\FF(n,k-1),
\end{equation}
The operations are applied from right to left, the MSB ($n$) is designated to attain the highest bias, $3BC(k)$ is the 3-bit compression of bits $k$, $k-1$ and $k-2$, implemented by the gate 
$\ket{100}\leftrightarrow\ket{011}$ (the MSB is on the left), and $\FF(n,2)$ is a RESET step on the two LSBs (bits~1 and~2). 

%
%
In the Fibonacci algorithm, the number $m_{n,k}$
of iterations can be freely chosen.
In the $m$-Fibonacci algorithm~\cite{EMW11}, $m_{n,k}=m$ is fixed
(for all $n,k$). 
For example, with $m_{n,k}=2$, the algorithm 2-Fib (or 2Fib) 
acting on $n$ spins, has this 
subroutine in it:
$\FF(n,4)= \FF(n,3) 3BC(4) \FF(n,3) 3BC(4) \FF(n,3)$. After another step in
the recursion, we get
\begin{eqnarray*}\FF(n,4)&=& 
\FF(n,2) 3BC(3) \FF(n,2) 3BC(3) \FF(n,2)
3BC(4) \\
&\cdot&\FF(n,2) 3BC(3) \FF(n,2) 3BC(3) \FF(n,2)
3BC(4) \\
&\cdot&\FF(n,2) 3BC(3) \FF(n,2) 3BC(3) \FF(n,2).
\end{eqnarray*}
For infinite $m$ the resulting biases of cooled spins are exactly the
Fibonacci series $\{\ldots,13,8,5,3,2,1,1\}$. 

%
%
Regarding $\FF(n,2)$ we consider (in all the Fibonacci algorithms) two cases:
\begin{itemize}
\item Both LSBs are reset bits: $\FF(n,2)=WAIT$
\item Only bit~1 is a reset bit: $\FF(n,2)=\FF(n,1)2BC(2)\FF(n,1)$, where $2BC$ ($\ket{10}\leftrightarrow\ket{01}$) is SWAP and $\FF(n,1)=WAIT$.
\end{itemize}
Clearly the second case simply doubles the number of reset steps.

%
%
In the algorithm $delta$-Fibonacci, 
the number of iterations~$m_{n,k}$ is chosen so that the bias of bit $k$ 
will become at least $F_{k}\left(1-\delta^{n-(k-1)}\right)\eps_0$, 
where $F_k$ is the $k\th$ number in the Fibonacci sequence. 
For  $\delta=\textonehalf$ the required number of iterations  $m_{n,k}$ is bound from above~\cite{SMW07,EMW11} by $n-k+2$. 
The MSB will then attain a bias increase factor of at least $F_n/2$, 
where the runtime of the algorithm will be bound by  $n!$~\cite{SMW07,EMW11}.

%
%
For example if we take 4 bits, and $\delta=\frac{1}{2}$,
the goal will be $\left\{1.5,1.5,\frac{7}{8},\frac{15}{16}\right\}$.  
The algorithm $\FF(4,3)$  will be as follows:
\begin{eqnarray}\label{eq:FF(4,3)}
\{0,0,0,0\}  \xrightarrow{\FF(4,2)}  \{0,0,1,1\}&\xrightarrow{\FF(4,2)3BC(3)}&
 \{0,1,1,1\}\\
 &\xrightarrow{\FF(4,2)3BC(3)}&
 \{0,1.5,1,1\}\nonumber\\
 &\xrightarrow{\FF(4,2)3BC(3)}&
 \left\{0,1.75,1,1\right\}\nonumber\\
& \ldots & \ \nonumber
\end{eqnarray}
Notice that the goal has been reached already at the second iteration, 
$m_{4,3}$=2, if we skip the third iteration,  
then $\FF(4,4)$ with $m_{4,4}=2$ will be as follows:
\begin{eqnarray}\label{eq:FF(4,4)}
\{0,0,0,0\}  \xrightarrow{\FF(4,3)}  \{0,1.5,1,1\}&\xrightarrow{\FF(4,3)3BC(4)}&
 \{1.25,1.5,1,1\}\\
 &\xrightarrow{\FF(4,3)3BC(4)}&
 \left\{1\frac{7}{8},1.5,1,1\right\}\nonumber
 \end{eqnarray}
Which is already beyond the goal.  
If we choose $m_{4,3}=3$ as the bound allows, then $\FF(4,4)$ 
with $m_{4,4}=2$ will be as follows:
\begin{eqnarray}\label{eq:FF(4,4) bound}
\{0,0,0,0\}  \xrightarrow{\FF(4,3)}  \{0,1.75,1,1\}&\xrightarrow{\FF(4,3)3BC(4)}&
 \left\{1\frac{3}{8},1.75,1,1\right\}\\
 &\xrightarrow{\FF(4,3)3BC(4)}&
 \left\{2\frac{1}{16},1.75,1,1\right\}\nonumber
 \end{eqnarray}

%
%
For the case of 4 bits it is also possible to run Tribonacci: \\
\emph{Tribonacci: Run $\TT(n,n)$}
\begin{equation}\label{eq:Tribonacci-definition}
	\TT(n , k)=\left[\TT(n,k-1)\; 4BC(k)\right]^{m_{n,k}}\TT(n,k-1),
\end{equation}
Where $4BC$ is implemented by $\ket{1000}\leftrightarrow\ket{0111}$, $\left(\textrm{which for 
biases $\ll1$ achieves }
\eps_4\leftarrow\frac{3\eps_4+\eps_3+\eps_2+\eps_1}{4}\right)$, and
$\TT(n,3)=\FF(n,3)$ 
is Fibonacci on the three LSBs, run enough times to reach the desirable bias.
In the case of 4 bits, the algorithm is simply $4BC(4)$ and $F(4,3)$ 
iteratively and the 4-term Tribonacci sequence \{4,2,1,1\} 
is asymptotically approached. Similar to Fibonacci, we can choose a goal of 
$T_{k}(1-\delta^{n-(k-1)})\eps_0$ for each bit $k$, 
where $T_k$ is the $k\th$ Tribonacci number. So the
goal for 4 spins will be $\left\{2,1.5,\frac{7}{8},\frac{15}{16}\right\}$.
The values of $m_{n,k}$ satisfying this goal, with $\delta=\frac{1}{2}$ can be 
chosen to be $m_{4,3}=2$ and $m_{4,4}=3$. 
\begin{eqnarray}
\{0,0,0,0\}\xrightarrow{\FF(4,3)}\{0,1.5,1,1\}&\xrightarrow{\FF(4,3)4BC(4)}&
\left\{\frac{7}{8},1.5,1,1\right\}\label{eq:TT(4,4),m=2}\\
&\xrightarrow{\FF(4,3)4BC(4)}&
\left\{1\frac{17}{32},1.5,1,1\right\}\nonumber\\
&\xrightarrow{\FF(4,3)4BC(4)}&
\left\{2\frac{3}{128},1.5,1,1\right\}\nonumber
\end{eqnarray}
We can also choose $m_{4,3}=3$ and $m_{4,4}=2$
in order to compare it to our calculations of $\delta$-Fibonacci:
\begin{eqnarray}
\{0,0,0,0\}\xrightarrow{\FF(4,3)}\{0,1.75,1,1\}&\xrightarrow{\FF(4,3)4BC(4)}&
\left\{\frac{15}{16},1.75,1,1\right\}\label{eq:TT(4,4),m=3}\\
&\xrightarrow{\FF(4,3)4BC(4)}&
\left\{1\frac{41}{64},1.75,1,1\right\}\nonumber\\
&\xrightarrow{\FF(4,3)4BC(4)}&
\left\{2\frac{43}{256},1.75,1,1\right\}\nonumber
\end{eqnarray}
where we provide also $m_{4,4}=3$ for comparison.

%
%
We see that for  the same $m_{n,k}$ values used for $\delta$-Fibonacci, the Tribonacci algorithm not only fails to reach its goal, it even 
yields lower values than Fibonacci algorithm.
Clearly, something is wrong in the definition of the Tribonacci algorithm as is
given in~\cite{EFMW06,EFMW07}. This problem becomes even more severe when
looking at high-term bonacci algorithms of~\cite{EFMW06,EFMW07}, such as
4-term-bonacci, etc., and finally k-term-bonacci and All-bonacci.
In order to identify the source of the problem it is best to look back into the
Fibonacci algorithm.

%
%
We can easily identify that the problem exists also in the Fibonacci algorithm:
We notice that at the first iteration of $\FF(4, 4)$ in Eq.~\ref{eq:FF(4,4)}, replacing the 3BC with $2BC(4)$ will attribute bit 4 with a higher bias \begin{eqnarray}\label{eq:new-FF(4,4)}
\{0,0,0,0\}  \xrightarrow{\FF(4,3)}  \{0,1.5,1,1\}&\xrightarrow{\FF(4,3)2BC(4)}&
 \{1.5,1.5,1,1\}\\
 &\xrightarrow{\FF(4,3)3BC(4)}&
 \{2,1.5,1,1\}\nonumber
\end{eqnarray}
Following Eq.~\ref{eq:FF(4,4) bound} and setting $m_{4,3}=3$, we will achieve an even higher bias
\begin{eqnarray}\label{eq:new-FF(4,4) bound}
\{0,0,0,0\}  \xrightarrow{\FF(4,3)}  \{0,1.75,1,1\}&\xrightarrow{\FF(4,3)2BC(4)}&
 \{1.75,1.75,1,1\}\\
 &\xrightarrow{\FF(4,3)3BC(4)}&
 \{2.25,1.75,1,1\}\nonumber
 \end{eqnarray}
Notice that in both Eqs.~\ref{eq:new-FF(4,4)} and~\ref{eq:new-FF(4,4) bound} 
the goal has been achieved already  after a single iteration. 
When running the Fibonacci algorithm on more than 3 bits  
replacing $3BC(k)$ by $2BC(k)$ in the first 
iteration of each recursive level will increase the final bias,
and the larger $k$ is --- the larger is the gain relative to the original
Fibonacci algorithm. 
Alternatively the same bias goal of $F_{k}\left(1-\delta^{n-(k-1)}\right)\eps_0$ can be 
achieved with less iterations. We name this version of the 
algorithm \emph{new-Fibonacci}. 

\begin{figure}[h]
\label{Figures} 
\includegraphics[width=0.5\columnwidth]{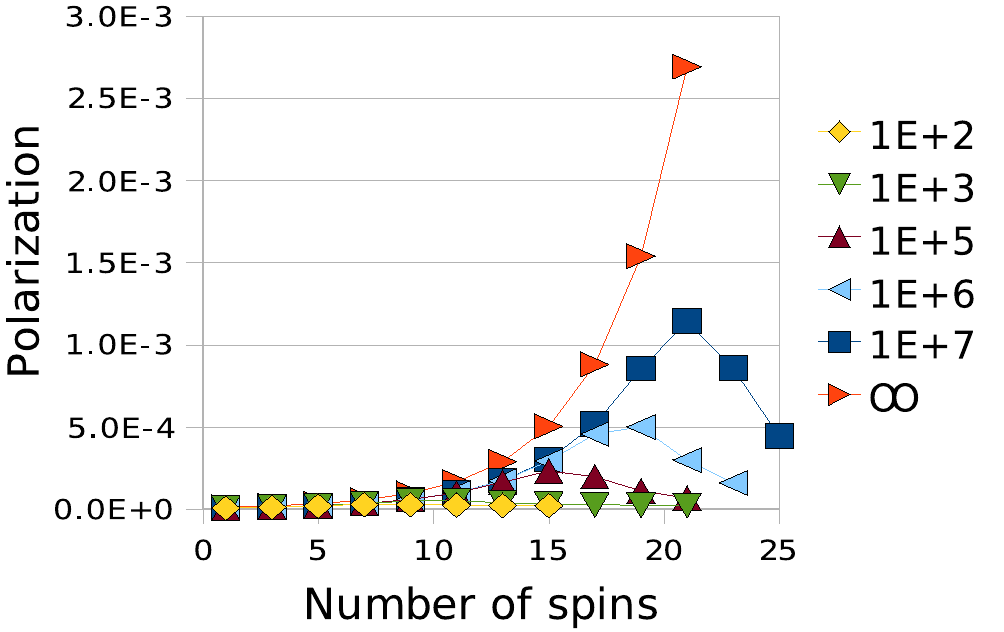}
\includegraphics[width=0.5\columnwidth]{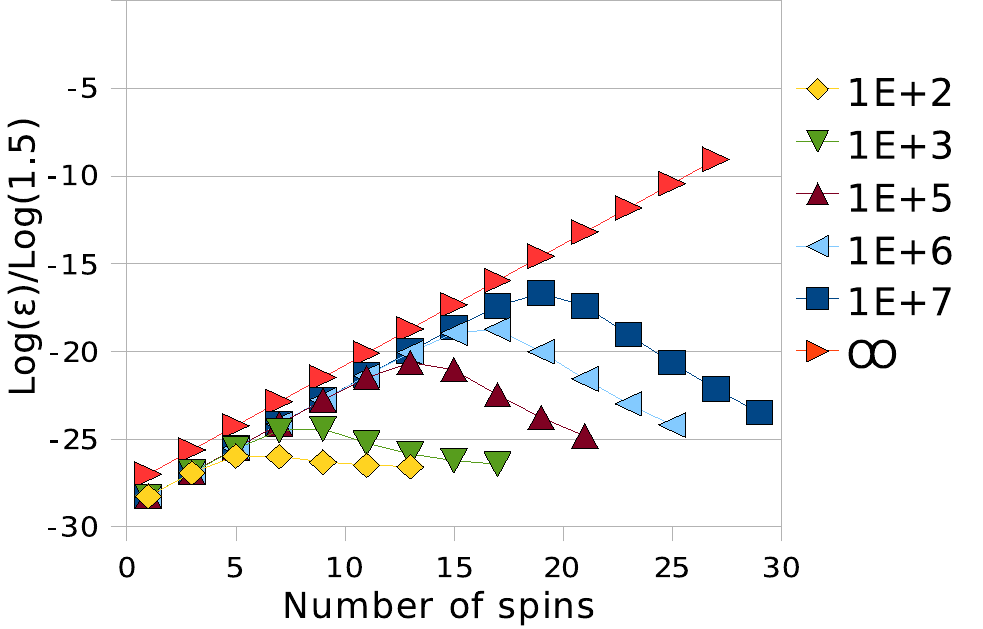}
\caption{Simulation results: the diagram on the left shows the maximal bias achievable by 2PAC on spin systems with various $\RR$ values ranging from 100, the scale which may be attainable in liquid-state NMR, to $10^6$ attainable in solid-state NMR using spin-diffusion as the reset mechanism; where the red graph is the ideal case of $\RR = \infty$. The equilibrium bias is set to $\eps_0=10^{-5}$. We see that even for the enourmous $\RR=10^7$, bias increase by 2PAC is not possible for more than two orders of magnitude and for no more than 20 spins. The diagram on the right is the same diagram where the $y$ axis is in logarithmic scale, which shows us that the cooling is exponential until reaching the highest achievable bias, even for the smallest $\RR$ simulated.}
\end{figure}

%
%
An interesting example is when we choose a constant number of 
iterations, $\forall\{ n,k\},\, m_{n,k}= 2$. 
In this case the original Fibonacci algorithm will become the algorithm 2Fib,  
as earlier mentioned,
and the (better) new Fibonacci, will become new-2Fib, identical to an  
algorithm named PAC3~\cite{EMW11}. 
So from what we learn here about old vs.\ new Fibonacci 
we conclude that PAC3 always outperforms 2Fib.

%
%
Similar to new-Fibonacci, we also define new-Tribonacci: 
it is wiser to use $2BC(k)$ instead of $4BC(k)$ in the 
first iteration of each recursive level, and in the next iteration(s) 
$3BC(k)$ will 
give the highest bias to bit $k$. Choosing $m_{4,3}=2$ we get
\begin{eqnarray}\label{eq:new-T(4,4)}
\{0,0,0,0\}\xrightarrow{\FF(4,3)}\{0,1.5,1,1\}&\xrightarrow{\FF(4,3)2BC(4)}&
\left\{1.5,1.5,1,1\right\}\label{eq:new-TT(4,4)m=2}\\
&\xrightarrow{\FF(4,3)3BC(4)}&
\left\{2,1.5,1,1\right\}\nonumber\\
&\xrightarrow{\FF(4,3)4BC(4)}&
\left\{2.5,1.5,1,1\right\}.\nonumber
\end{eqnarray}
And choosing $m_{4,3}=3$ we get:
 \begin{eqnarray}\label{eq:new-T(4,4) bound}
\{0,0,0,0\}\xrightarrow{\FF(4,3)}\{0,1.75,1,1\}&\xrightarrow{\FF(4,3)2BC(4)}&
\left\{1.75,1.75,1,1\right\}\label{eq:new-TT(4,4)m=3}\\
&\xrightarrow{\FF(4,3)3BC(4) or 4BC(4) }&
\left\{2.25,1.75,1,1\right\}\nonumber\\
&\xrightarrow{\FF(4,3)4BC(4)}&
\left\{2\frac{5}{8},1.5,1,1\right\}.\nonumber
\end{eqnarray}
  Notice that for both Eqs.~\ref{eq:new-T(4,4)} and~\ref{eq:new-T(4,4) bound} 
the goal has already been reached by the 2\nd iteration, before we even used 
a 4BC gate. 
This potentially reduces the time 
complexity for running Tribonacci or All-bonacci~\cite{EFMW06,EMW11} on larger 
spin systems. 

%
%
We define new-Tribonacci to be better then new-Fibonacci,
by chosing the optimal compression gate (2BC, 3BC, or 4BC) at each step.
Similarly, new-$k$-bonacci is therefore better than new-\{k-1\}-bonacci,
by definition.

%
%
It is interesting to research whether the new-$\delta$-bonacci algorithms 
can converge to the goal as defined earlier within a reasonable time. 
We know that the time 
complexity of the old version of $\delta$-fibonacci is $O(n!)$, 
hence, by definition, any new-bonacci will surely reach the Fibonacci goal,
but what is 
the complexity of new-$\delta$-Tribonacci for reaching the Tribonacci goal?
And similarly, for the other k-bonacci algorithms? And how significant 
is the advantage of the new versions of the algorithms? These questions are left
open for future research.

%
%
Another open research question arises from the redundancy in
the algorithms. E.g., after the first $2BC(4)$ 
step in new-Tribonacci, the next subroutine 
is $\FF(4,3)$ where the first step is a redundant reset $\FF(4,2)$
because the $2BC(4)$ step did not destroy the two reset spins anyhow.
So clearly it is interesting in practice (and maybe also possible in theory) to
optimize the algorithm further by removing such redundant steps.

\section{Accounting for relaxation}\label{sec:account-for-T1}
%
%
The definition of mPAC is
\begin{eqnarray} \label{eq:mPAC-definition}
M_{j}(k) &=& \left[\BB(k) M_{j-1}(k-2)\; PT(k-2\rightarrow k-1)\right.\\
       & & \left.M_{j-1}(k-2) \right]^{m} PT(k-2\rightarrow k)\; M_{j-1}(k-2).
       \nonumber
\end{eqnarray}
The polarization transfer (PT) moves the bias of the first spin onto the second
spin, such that $PT(A\rightarrow B)$ assigns 
to spin B the bias of spin A. 
For the purpose of this work, whenever PT is used,
it is identical to SWAP as we do not care 
about the polarization of spin $A$
after the PT or SWAP operation; and the advantage of PT is that
it is easier to implement in practice.

For example, using the genral formula above, 
2PAC on 3 spins is
 defined as
\begin{eqnarray} \label{eq:2PAC-3spins}
M_{1}(3) &=& \left[3BC(3) M_{0}(1)\; PT(1\rightarrow 2)\; M_{0}(1) \right]^{2} PT(1\rightarrow 3)\; M_{0}(1),
\end{eqnarray}
where $M_0(1)$ is a reset step of the reset bit (spin 1).
For 5 spins 2PAC is defined as follows:
\begin{eqnarray} \label{eq:2PAC-5spins}
M_{2}(5) &=& \left[3BC(5) M_{1}(3)\; PT(3\rightarrow 4)\; M_{1}(3) \right]^{2} PT(3\rightarrow 5)\; M_{1}(3).
\end{eqnarray}
The algorithm under this definition cools only the MSB. In order to cool all the bits from 1 to k we add the cooling of all less significant bits in decreasing order. For example
\begin{eqnarray}
M_{all}(3) &=& M_{all}(1)\; PT(1\rightarrow 2)\; M_0(1)\;M_1(3)\\
M_{all}(5) &=& M_{all}(3)\; PT(3\rightarrow 4)\; M_1(3)\;M_2(5)\\
M_{all}(k) &=& M_{all}(k-2)PT(k-2\rightarrow k-1)\; M_{j-1}(k-2)\;M_j(k),
\end{eqnarray}
where $M_{all}(1)=M_0(1)=WAIT$. The simulation results we show here are of $M_{all}(7)$.

We shall now deal with finite \RR, and analyze 2PAC, assuming one reset spin,
 for several values of \RR.
Let $\Twait = d\cdot\Toner$ be the duration of each reset step,
namely, we decide here that the time we WAIT and do nothing between
computing steps of the algorithm, is always the same.
When $d\approx 5$, the polarization bias 
of the reset spin 
approaches its equilibrium value $\eps_0$ by the end of each reset step.
We also want $\Twait\ll\Tonec$ in order to enable many reset steps in our 
algorithm. 
Ignoring the duration of unitary transformations, if an algorithm has 
$ N_r$ reset steps, each requiring a waiting time \Twait, then the runtime of
the algorithm is 
$\Trun = \Twait\cdot N_r = d\cdot N_r\cdot\Toner.$
For given $\Twait$ and $\Tonec$, it is desired that 
the runtime 
satisfies $\Trun = \Twait\cdot N_r \ll \Tonec$, 
hence we get a condition on the number of reset steps $ N_r$.
When the runtime satisfies $\Trun\ll\Tonec$, and $d=5$, the ideal results 
obtained in~\cite{EMW11} still apply with only minor corrections.
However, when $\Tonec/\Trun$ is in the range 
of about $0.1$ to $10$, namely, 
$\RR/(d\cdot N_r)$ is between $0.1$ and $10$, 
the deviation from the ideal result is significant, while for 
$\RR/(d\cdot N_r) > 10$ the deviation is still small.
Although this is redundant, it is still 
useful to define another dimensionless parameter $\QQ=\Tonec/\Twait$, 
such that $1/\QQ$ provides the extent by which the computation spins relax during
 one WAIT period.

%
%
In order to run a long algorithm (with many reset steps) and still reach
near-ideal cooling, we would want the durations to satisfy 
$\Tonec \gg \Trun \gg \Twait  \gg \Toner$. 
It is thus natural to define another (redundant) dimensionless parameter, 
$\DD=\Tonec/\Trun$ (hence $\DD = \QQ/ N_r$), and now 
$\Tonec = \DD \Trun = \DD  N_r  \Twait = \DD  N_r d \Toner$,
hence $\RR=\DD N_r d$. 
Note that as long as $\DD \gg 1$ (namely, $\QQ \gg  N_r$) the computation
spins barely relax, and as long as $d \gg 1$ the reset spin
regains a bias very close to its equilibrium bias.
 
%
%
Using a constant \Twait in our calculations allows a major simplification:
We can simply replace the equilibrium reset-spin bias by
$(1-e^{-d}) \eps_0,$ 
assuming the reset spin has no polarization prior to each reset step.
For each computation spin, one might expect to replace
the ideal final bias, $\eps_f$ by  
$e^{-( N_r/\QQ)} \eps_{f}$ (namely, $e^{-(1/\DD)} \eps_{f}$), 
which would be true if that spin  
got its final optimal bias at the beginning of the relaxation
process. However, the bias is gradually increased by
portions that are subject to less relaxation, 
hence the above result provides
a lower bound on the final polarization.

%
%
For $n$ spins the calculations are actually rather cumbersome, 
as there are $2^n$ terms in the diagonal density matrix, at 
each step.  Taking into account finite $d$ and $\QQ$ becomes non-trivial,
hence we take here an \emph{extended-Markovian} approach: after each 
reset step, not only does the environment ``forget'' 
any remaining correlations with
our system, but also the reset spin is considered as a separate
``environment'' and becomes uncorrelated with the computation 
spins. 
For $\RR\rightarrow\infty$, calculations using this model become identical to 
previous calculations done for ideal theoretical
AC~\cite{BMRVV02,FLMR04,SMW05,SMW07,EFMW06,EFMW07,HRM07,Kaye07,EMW11},  
if both $d$ and $\QQ$ go to infinity.  
When $d$ is finite, and we use this model,  
the bias of the reset spin evolves 
from some initial value $\eps$ into $(\eps -\eps_0)e^{-d}+\eps_0$. Already 
for $d=5$ the extended-Markovian approach seems reasonable, as can be verified
(numerically) 
for small values of $n$. However, it might be much less accurate for larger 
values of $n$, if compared with a model taking all correlations into
account.
The extended Markovian model influences the cooling 
process by directly removing only the correlations of the
 reset spin with the computer spins. However, for mPAC, this then 
indirectly removes undesired correlations among the computer spins as well.

%
%
Using this model, we simulated 2PAC with 7 spins, 
 starting from $\eps_0 \ll 1,$ with reset steps satisfying  
$d=5.$ Cooling \underline{all} the spins,
see~\cite{EMW11} section IV, requires $ N_r = 187$ steps, and
ideally attributes the most significant bit (MSB) with a final bias of $5.36$
 (in units of $\eps_0$), see Table II in~\cite{EMW11}.
This cooling level remains, as long as $\DD \gg 1$; When $\RR=10000$
 (i.e. $\QQ=2000$), the MSB attains a final bias of $5.11$.
When $\DD \sim 1$, the cooling level is significantly reduced;
$\RR=1000$ (namely $\QQ= 200$) leads to a final bias of $3.63.$
Finally, when $\DD \ll 1,$ the cooling becomes quite negligible;
When $\RR=100$ (namely $\QQ= 20$), the final bias is only $1.07$. 
Better results could be achieved if the two LSBs are reset bits.
Another way to reach better results is to cool only the MSB.
 Notably, $\RR \sim 10000$ was
 achieved in solid-state NMR, using spin-diffusion for rapid
 repolarization~\cite{BMR+05, RMBL08}, but even $\RR=100$ seems beyond
the capability of current liquid-state NMR\@.

The formal proof that the consequence of the extended Markovian model  
for the mPAC algorithm is the removal of all the spin correlations is not
given here with full details,
but let us consider
an illustrative example of using 2PAC on the four LSB of a spin system with four or more spins,
where only $A$ is a reset spin.
Suppose that at some stage in the process of cooling, the 3~spins  
have a known bias configuration $\{\eps_C,\eps_B,\eps_A\}$, such that the spins
are potentially correlated among them and/or with other spins.
In the extended Markovian model,
cooling of these 3~spins, by repeated reset steps and SWAPs, removes their
correlations, as we now show.
We start after spin $A$ had been reset to some near-equilibrium --- 
 after a WAIT of a duration of $5\cdot T_1(A)$, the bias of $A$ will 
be $\eps_r=(\eps_A-\eps_0)e^{-5}+\eps_0$, 
with no remaining correlations with the other spins, due to the
extended-Markovian assumption. 
The reset of the three spins now goes as follows:
\begin{equation}
\left\{\underline{\eps_C},\underline{\eps_B}, \eps_r\right\} \XRA{A\leftrightarrow C}
\left\{\eps_r,\underline{\eps_B},\underline{\eps_C}\right\} \XRA{WAIT}
\left\{\eps_{r'},\underline{\eps_B},\eps_r\right\}
\XRA{A\leftrightarrow B}
\left\{\eps_{r'},\eps_r,\underline{\eps_B}\right\}
\XRA{WAIT}
\left\{\eps_{r''},\eps_{r'},\eps_r\right\}, 
\end{equation}
where
$\eps_{r'}=\left(\eps_r-\eps_0\right)e^{-\frac{T_{WAIT}}{T_1(comp)}}+\eps_0$,
and $\eps_{r''}$ is similar but with $2T_{WAIT}$, and where we underline each 
spin when it is still potentially correlated.
We continue according to $m$PAC 
(see Eq.~\ref{eq:mPAC-definition}) for $m=2$, applied onto 4~spins, $DCBA$, 
where only $A$ is a reset spin and the bias configuration is 
$\left\{\eps_D,\underline{\eps_C},\underline{\eps_B},\underline{\eps_A}\right\}$, as is typical after a 3BC on bits $CBA$. Here 
$\eps_D$ is unknown and $D$ is not correlated to any other spin(s). We underline the potentially correlated spins. 
 The next step of 2PAC is using SWAP steps between $D$ and the cooled $C$
leading to the configuration$\left\{\underline{\eps_C},\eps_D,\underline{\eps_B},\underline{\eps_A}\right\}$
 After a WAIT of a duration of $5\cdot T_1(A)$, the bias of $A$ will 
be $\eps_r=(\eps_A-\eps_0)e^{-5}+\eps_0$ with no remaining correlations with the other spins\footnote{In the simulation we chose $\eps_A=0$, and so
$\eps_r=(1-e^{-5})\eps_0$. The computation spins also relax during the WAIT, but we ignore this here, dealing only
with the issue of removal of correlations. In the simulation we took all relaxations into consideration.}. The extended Markov assumption means that spin $A$ will now be uncorrelated to the rest of the spins on the molecule, and therefore  in a tensor product state with all other spins. The following steps of 2PAC are:
\begin{equation}
\left\{\underline{\eps_C},\eps_D,\underline{\eps_B}, \eps_r\right\} \XRA{A\leftrightarrow C}
\left\{\underline{\eps_C},\eps_r,\underline{\eps_B},\eps_D\right\} \XRA{WAIT}
\left\{\underline{\eps_C},\eps_{r},\underline{\eps_B},\eps_r\right\}
\XRA{A\leftrightarrow B}
\left\{\underline{\eps_C},\eps_{r},\eps_r,\underline{\eps_B}\right\}
\XRA{WAIT}
\left\{\eps_C,\eps_{r},\eps_{r},\eps_r\right\}.
\end{equation}
It is easily possible to prove now, by induction, 
that for any number of spins, correlations are completely removed 
prior to all compression steps.

\section{Lower bounds on required reset steps}
\label{sec:AC-bound-resets}
%
%
The previous section dealt with a specific
practicable algorithm. Could much
better results be obtained using a much more efficient algorithm,
or even the optimal algorithm (PPA)?

AC relies on thermalization (or other means of repolarization) for reset
operations following selective PT from a reset 
spin to the target spin or following polarization compression. 
Ideally, optimal AC asymptotically cools one computation spin in an $n$-spin
 system (with one reset spin) by a factor of
$2^{n-2},$ when the final polarization is $\lesssim 1\%$. The
 optimal PPA algorithm was proven to approach this limit, 
and to require a minimal number of reset steps at each
 stage~\cite{SMW05,SMW07}. 
As in the previous section, also here we neglect the time of the unitary
evolution in between reset steps.
The use of an extended-Markov model cannot be justified for the PPA as 
correlations among computation spins are not removed by reset steps followed by 
SWAPs, etc.  Thus, we analyze here the ideal case with infinite $\QQ$. 
If $d$ is infinite then this is exactly the ideal model, and if $d$ is finite
then only the bias $\eps$ should be modified accordingly.
Since we focus on infinite $\QQ$, the limitations here are not as novel
as the ones discussed in the above section, but are simple consequences of
our earlier results together with Schulman~\cite{SMW05,SMW07}. 
Such results, when
written the way they appear here, simplify the comparison of AC with other methods
used to improve the SNR in magnetic resonance spectroscopy. 

%
%
We now present 
lower bounds (based on entropy considerations) 
on the number of reset steps required by 
any spin-cooling algorithm~\cite{SMW05,SMW07}.
\begin{thm}\label{thm:AC-resets}
Given an $n$-spin system with a single reset spin, and $n-1$
computation spins, where all spins are in
a completely mixed state and the polarization bias of the reset spin 
after a reset step is $\eps
 \ll 1,$ the lower bound on the number of reset steps required to cool one
 computation spin to $k\eps \ll 1$ is $k^2.$
\end{thm}
\proof ~

Our goal is to cool one spin to $k\eps,$ whereby the total entropy
of the spin system, $H_{init} = n,$ would decrease to $H_{fin} \le 
 n-(k\eps)^2/\ln{4} = n-k^2\eps^2/\ln{4};$ the equality is obtained when
the other $n-1$ spins remain completely mixed (no other spin is cooled).
The entropy removed from the entire spin system is therefore
 $\Delta H = H_{init} - H_{fin} \ge \left(k^2\right) \eps^2/\ln{4}.$
Each reset step reduces the entropy by at most~\cite{POTENT} the information
content of a single reset-spin at thermal equilibrium $I_1= \eps^2/\ln{4}$ (when
 the reset spin goes from the
 completely mixed state to equilibrium). Therefore the lower bound on the number
 of reset steps is given by: $\frac{{\Delta H}}{I_1} \ge
 \frac{\left(k^2\right) \eps^2/\ln{4}}{\eps^2/\ln{4}} = k^2.$
\qed\\

>From this theorem it follows that for optimal 
cooling~\cite{SMW05}, where $k = 2^{n-2},$ the lower bound on the number of
 required reset steps is about $k^2$ already for
 small spin systems, since $k^2 = 2^{2(n-2)} \gg n$ is satisfied already for $n
 \gtrsim 6.$ It is therefore much more efficient 
(in terms of average entropy reduction per reset step) 
to stop the PPA after much fewer reset steps.

\section{Prospects of algorithmic cooling and heat-bath cooling for NMR
 spectroscopy}\label{sec:prospects}
We investigated the feasibility of applying the heat-bath cooling (HBC)
 experiments, as well as algorithmic cooling (AC), to improve
NMR spectroscopy. We emphasize in vivo \isotope{C}{13} spectroscopy
 (13C-MRS) in the brain, where \isotope{C}{13}-labeled metabolites are observed over
 extended periods of time (up to several hours).
One of the main targets of in vivo brain spectroscopy has been glutamate (Glu),
a major excitatory neurotransmitter, to which we applied HBC~\cite{EGMW11}. Despite focusing on 13C-MRS, the results here,
 especially those in section~\ref{sec:AC-sensitivity}, are more general. 

\subsection{Carbon-based brain spectroscopy}\label{sec:13C-MRS-brain}
%
%
The application of NMR spectroscopy in vivo, for non-invasive study
of metabolic processes in living organisms, is called \emph{magnetic resonance spectroscopy
 (MRS)}. Proton (\isotope{H}{1}) MRS is applied clinically since the late 1980s, and
 was approved by the US Food and Drug Administration (FDA) about a decade ago; 
It is commonly performed on conventional MR Imaging (MRI) systems.
In the brain, proton MRS 
 measures about 35 metabolites,
 a small portion of the estimated 2,000-20,000 metabolites~\cite{MSLRR10}.
 Localization to a specific volume in the brain allows to focus on the
 metabolism in a small region of interest, such as a malignant
 tissue. Proton MRS is also applied clinically for diagnosis of metabolic
 disorders, where abnormal enzymatic activity produces unusual metabolite
concentrations~\cite{vanderGraaf10}. MRS of the living brain provides a
 ``virtual biopsy''~\cite{MSLRR10}, revealing information not
accessible by other means.

%
%
One main goal of (brain) MRS is to provide useful diagnostic information for
clinical investigation; for this purpose, the steady-state concentration
 measurements obtained by \emph{static analysis} with proton MRS are quite
limited~\cite{RLH+03}.
 While proton MRS is very common, carbon is the
 ideal element for following metabolic processes; it is present in most
 metabolites and may be labeled by enriching the stable isotope \isotope{C}{13}
from its natural abundance of $1.1\%$ up to $\sim 99\%.$ The label can be
followed dynamically, over extended periods of time, with negligible background,
 since the major isotope (\isotope{C}{12}) lacks magnetic moment. In contrast,
hydrogen-based MRS is not useful for following dynamic processes,
since both stable isotopes are NMR-active, and the much less abundant \isotope{H}{2}
isotope (spin-1) has very low sensitivity and small chemical shifts (about six
times smaller than for \isotope{H}{1}).

%
%
13C-MRS is developed since the 1970s and was shown in the early 1990s to
detect labeled glutamate (Glu) in the human brain~\cite{GAC+03} following
 infusion of \isotope{C}{13}-labeled glucose. It provides enhanced resolution
 over proton MRS, due to the much wider (by over 20-fold) chemical shift range
 of \isotope{C}{13}~\cite{RC05, DRB11}. However, its SNR is significantly lower than
proton MRS, such that many scans are generally needed for a clinically-useful
 diagnosis. 
While technically challenging, 13C-MRS has been established as the only
 noninvasive method for measuring neurotransmission by Glu and the energetics of
 specific cells in the human brain~\cite{RDDMB11}. It has provided rates for
specific metabolic routes, such as neurotransmitter cycling
 (Glu/glutamine)~\cite{RLH+03}, allowed to establish the energy cost of brain
 function, and provided various other highly important information (reviewed in 
Ref~\cite{RDDMB11}). Recently, 13C-MRS also provided exchange rates for
 individual enzymes~\cite{XS09}, using saturation transfer.

%
%
13C-MRS supports \emph{personalized medicine}\footnote{People now consider
the four ``P''s:
 predictive, personalized, preventive, participatory~\cite{MSLRR10}.}
and may provide initial diagnosis, monitor treatment efficacy, and
 determine optimal treatment and dosage~\cite{MSLRR10}. In contrast, proton MRS is not useful for in vivo drug monitoring, since contrast agents either do not enter the brain, or they interact with tumors differently than the drug.
While still pending FDA approval for routine clinical usage,
13C-MRS is showing promising clinical potential; recently, by acquiring
spectra briefly at steady state after infusion of labeled acetate, 13C-MRS
 enabled rapid assessment of inflammation in glia cells, indicative of
 neurodegeneration~\cite{STHR10}. More recently, this protocol was applied 
to patients of Alzheimer's disease and showed an increase in glial 
metabolism, which was correlated with cognitive decline~\cite{STHR11}.
Several major challenges remain in the path of clinical 13C-MRS, 
including~(mainly) low SNR, which requires acquisition of relatively large 
volumes~(several mL), and also tissue heating and prolonged infusion 
periods~(typically 2 hours)~\cite{RDDMB11}.

%
%
By employing hyperpolarized substrates, the SNR of 13C-MRS can be dramatically
improved, enabling metabolic imaging~\cite{KVB+11}. Human clinical trials were recently conducted and showed promising results
for prostate cancer using dissolution dynamic nuclear polarization with 
hyperpolarized \isotope{C}{13}-pyruvate~\cite{NKV+13}.
Since the hyperpolarization decays after few minutes at most, applicability to
 brain spectroscopy is limited~\cite{RBW+10}.

\subsection{AC for \isotope{\mathbf C}{\mathbf 13} brain spectroscopy}
%
%
AC requires metabolites where there is a string of qubits, with at least
one reset spin.
The portion of such metabolites may be increased by using a suitable
 \emph{isotopomer} (isomer with specific labeling). 
For example, fully-labeled glucose might be a better
substrate than the single or double labeled isotopomers. Recently,
 metabolism of different substrates, e.g. glucose and lactate, was
 detected simultaneously at high field (11.7T), based on different
 line-splitting patterns~\cite{XS11b}. 
Heteronuclear PT from \isotope{H}{1} to \isotope{C}{13}, which may be considered as
a relevant starting point for 
AC and HBC, was successfully implemented in the brain in a recent in vivo
 study of aging~\cite{BMG+10}.

%
%
%
Theoretically, for $n$ spins, AC was shown to produce exponential cooling of
 $\left(\frac{3}{2}\right)^{\lfloor(n-1)/2\rfloor}$ via the most basic 
 algorithm~\cite{FLMR04}, and up to $2^{n-2}$~\cite{SMW05,SMW07, EFMW06} 
via the optimal algorithms. However, in practice, cooling is limited by various physical constraints, which include experimental relaxation/repolarization rates
(as shown in section~\ref{sec:account-for-T1}), decoherence, imperfect pulses, 
and RF inhomogeneity. Moreover, even in the ideal case, we show in 
section~\ref{sec:AC-sensitivity} that, contrary to initial 
expectations~\cite{FLMR04,EFMW06}, for improving the SNR of a particular spin
using one reset spin, it is often more efficient to perform multiple
 repetitions of PT than to use AC\@. It may, however, still
 be advantageous to perform AC in several cases; 
There are three important areas where AC could contribute to NMR
 spectroscopy:
when \emph{multiple reset spins} are available (which we do not discuss here);
\emph{spectral editing} to resolve overlapping signals
 (section~\ref{sec:AC-spect-edit}); and reduction of the 
\emph{specific absorption rate (SAR)}, which is highly important for in vivo
carbon-based brain spectroscopy (section~\ref{sec:SAR-reduce}).
 
\subsection{Sensitivity enhancement - AC vs signal-averaging}
\label{sec:AC-sensitivity}
It is interesting to note that the use of $k^2$ reset steps as in 
theorem~\ref{thm:AC-resets}, is also highly relevant in a more conventional
approach in NMR: signal averaging. By averaging over $ N_r = k^2$ scans, after PT from a reset spin the SNR is also increased by a factor 
of~$k$, hence the advantage of AC seems to be lost. However, the significantly 
fewer scans required by AC might be advantageous in reducing the specific 
absorption rate (SAR).

%
%
AC relies on thermalization (or other means of repolarization) for reset
operations. It is therefore natural to compare its performance with
 multiscan-PT, a form of signal averaging that
 enhances the SNR by performing several scans, each following selective PT from
 a highly-polarized reset 
 spin to the target spin, with a delay between scans to allow the reset spin to
 repolarize. Here we ignore the high polarization of the reset spin, which for
 protons and carbons simply adds a multiplicative factor of four to both AC and
 signal averaging, and we focus on the short reset time of the reset spin.
In contrast to the conventional multiscan used in signal averaging,
the delay between consecutive scans in multiscan-PT is much shorter, as it is
 determined by the relaxation time of a fast-relaxing spin.
Ideally, optimal AC asymptotically cools one computation spin in an $n$-spin
 system, where one spin is a reset spin, by a factor of
$2^{n-2},$ up to final polarizations of about 1\%~\cite{SMW05}. The optimal
 \emph{partner pairing algorithm (PPA)} approaches this limit using a 
 minimal number of reset steps. In practice (e.g., if analyzed using the tools introduced in section~\ref{sec:account-for-T1}) it is less clear if AC will be sufficiently efficient relative to multiscan-PT.

\subsection{AC for reduction of SAR}\label{sec:SAR-reduce}
%
%
The specific absorption rate (SAR) measures the rate at which
electromagnetic (RF) radiation is absorbed by the body, reflecting the RF power
 applied during pulse transmission. SAR levels (in W/kg) are tissue dependent,
e.g. in the frontal lobe (around the eyes) the SAR is high due to limited
 circulation~\cite{RDDMB11}. 
Local SAR levels are obtained by numerical methods, as it is difficult to
 measure the relevant electrical fields or temperatures in vivo~\cite{DRB11}.
 Alternatively, a three-dimensional sample (phantom) is used, which mimics the
 target tissue~\cite{LZW+09}. For in vivo MRI and MRS, SAR levels are tightly
regulated; for the entire head, the maximum SAR is 3W/kg averaged over
 10 minutes, while for any other tissue, 8W/kg are allowed, averaged over 5
 minutes~\cite{DRB11}.

%
%
For 13C-MRS, \emph{proton decoupling}, where the protons are saturated with
RF irradiation, is commonly employed during data acquisition to considerably
simplify the spectrum (reduce the line splittings of each \isotope{C}{13} signal), thereby
increasing both SNR and resolution. For example, at 3T the signals of the
 carboxylic carbon \Cone\ of aspartic acid and Glu largely overlap without
decoupling, while they are well-resolved when decoupling is applied (see Fig. 4
 in ref~\cite{LZW+09}). 
Proton decoupling is typically applied over a prolonged period of time (100
\msec\ to few seconds), and therefore strongly contributes to the
 SAR~\cite{DUH06}. 
The large heteronuclear scalar coupling ($\sim 125$--$150~\Hz$) between alkyl
carbons (e.g., Glu \Cfour) and their attached protons presents a major problem
 for in vivo 13C-MRS~\cite{DRB11}, since effective decoupling requires very
high RF field strength ($\gamma B_2 \gg J_{CH},$ where $B_2$ is the decoupler
field)~\cite{LYS07}; In order to meet SAR limits, decoupling is generally
 limited to surface or half-volume coils rather than the state-of-the-art
 volume coils.
The problem is much worse at higher fields; the decoupling bandwidth 
grows linearly with magnetic field strength, and the SAR grows quadratically.
Already at 3T, SAR limitations considerably reduce the effectiveness of
WALTZ-4, the most common in vivo decoupling method that performs well at 
conditions of low RF homogeneity~\cite{LYS07,LZW+09}. Indeed, concerns were
raised that the inability to decouple effectively might offset the benefits of
 higher magnetic fields commercially available.

%
In 2007, a strategy was devised~\cite{LYS07} to overcome the limitation on
decoupling described above. By using [2-\isotope{C}{13}]glucose, rather
than the commonly used 1-labeled isotopomer, the label was found to be
incorporated in the carboxylic carbons (mainly Glu \Cfive, but also
Glu \Cone\ and glutamine \Cone\ and \Cfive~\cite{SRH+08}).
Even though no protons are directly attached, decoupling is still needed in most
 cases, although much lower power is sufficient.
The SNR of carboxylic/amide carbons is reduced due to their long \Tone\ and the
inability to perform PT (since protons are unavailable). However, the
 absence of interfering lipid signals opens the possibility of parallel imaging
 using multiple \isotope{C}{13} receivers and whole brain low power
 decoupling~\cite{XS11}. 
Further reduction in decoupling power was achieved by using stochastic
 decoupling, originated by Ernst, which is based on random noise~\cite{LYS07}. 
Such decoupling provided in 2008 acceptable SAR levels at 1.5T in the frontal
 lobe in human subjects, which is associated with memory and other higher
 functions~\cite{SRH+08}. 

%
We have seen that when only one reset spin is available, multiscan-PT
offers better SNR improvement than AC\@. 
However, several acquisition periods (one per scan) are required by
multiscan PT, while AC only employs a single acquisition period at the end of
 the cooling algorithm. Consequently, AC could considerably reduce the SAR\@.
 For example, [1,2-\multisotope{C}{13}{2}]Glu produced in the brain after 
administration of [1,2-\multisotope{C}{13}{2}]glucose (see Fig. 5 in Ref~\cite{RC05})
 can be cooled (ideally by a factor of 6) by PAC1 (HBC followed by 3B-Comp) with
 two reset steps. Multiscan PT can cool better (by a factor of $4\sqrt{3} \sim
 6.8$), however the two additional proton-decoupled acquisition periods would
 generate up to three times higher SAR\@.
 AC could therefore accommodate the higher power required for standard WALTZ
 decoupling; Alternatively, the reduced SAR could enable simultaneous
acquisition of multiple volume elements (voxels) in the brain, which could
allow to compare normal and abnormal brain regions~\cite{LZW+09} by permitting
 the use of RF-intensive volume coils. 

%
It is not yet clear whether the reduction in dissipated heat associated with AC
 would yield a comparative reduction in the SAR\@. The recent 3T study of 
the frontal lobe~\cite{LZW+09} acquired spectra every 4\sec\ (irradiating the
proton for $205~\msec$ in each scan);
during this small inter-scan delay, the slow heat dissipation in the frontal
 lobe might still limit SAR accumulation over successive scans.
The actual effect of AC on the SAR may be
 estimated using well-established numerical simulations that consider tissue
 properties (see Ref~\cite{LZW+09} and references therein). 

\subsection{AC for spectral editing}\label{sec:AC-spect-edit}
%
%
We consider \emph{spectral editing} in the broad sense as an operation
 (pulse sequence) that enhances a selected part of a complex NMR spectrum. 
In particular, spectral editing techniques (1D and 2D) are useful for
 selecting subspectra of isotopically-enriched metabolites, such as
 \isotope{C}{13} and \isotope{N}{15}-labeled isotopomers~\cite{FL08}.
A need for spectral editing is commonly encountered in vivo, 
where many metabolites are produced. In many cases, several
 \emph{isotopomers} (isomers labeled at different positions) of the same
 metabolite are formed, e.g. single-labeled, double-labeled, and triple-labeled
Glu obtained after administration of [1,6-\multisotope{C}{13}{2}]glucose~\cite{DUH06,
DRB11}. 

%
%
We~\cite{EGMW11} applied simultaneous HBC of the two backbone
 carbons of amino acids, \Cone\ and \Ctwo. Both labeled carbons of Glu and Gly
 were cooled simultaneously by truncated PT via environment thermalization (POTENT) pulse sequences~\cite{POTENT}, which 
consisted of a selective-reset (an HCC relay - from the proton to \Ctwo\ to 
\Cone, followed by a reset step) and a second PT from the \Alp\ protons to 
\Ctwo. This pulse sequence may be used for in~vivo 
spectral editing of glycine isotopomers, to selectively enhance the
 double-labeled isotopomer; both carbons would be significantly enhanced
 ($\sim 2.5$-fold), while for the single labeled
[2-\isotope{C}{13}]Gly isotopomer, only \Ctwo\ would be enhanced (for
 [1-\isotope{C}{13}]Gly, \Cone\ would be diminished due to its long
 \Tone)\footnote{An HCC relay which
enhances only \Cone\ is sufficient to reveal the presence of the double
 labeled Gly isotopomer.}.
For glutamate (and glutamine), similar spectral editing could selectively
enhance the 1,2-\multisotope{C}{13}{2} \emph{cumomer}\footnote{The concept of a 
cumulative isotopomer, or cumomer, was introduced about a decade ago to
analytically solve nonlinear isotopomer balance equations~\cite{WMI+99}.},
 which includes all nine
isotopomers where both \Cone\ and \Ctwo\ are labeled, from the other three
cumomers produced after infusion of labeled glucose~\cite{RC05,DUH06}.

%
%
The pulses applied to the protons are expected to
produce limited PT to \Cthree\ and \Cfour, since the refocusing delay of the
heteronuclear INEPTs is set\footnote{Setting
the delay to be $\frac{1}{2J}$ would entirely suppress PT from the methylene protons
if all J couplings were the same. In practice~\cite{DUH06,FL08}, the stronger
scalar coupling of \Ctwo\ ($\sim 145~\Hz$ vs $\sim 130~\Hz$) is expected to
allow limited PT to \Cthree\ and \Cfour\ (at about 30--40\% efficiency).} according to CH~\cite{BE80}.
A slight modification of POTENT could prevent any residual polarization
 enhancement on proton-bearing carbons other than \Ctwo, e.g. \Cthree\
 and \Cfour\ of Glu and glutamine. In the improved variant of POTENT, the
 hard pulses applied to the protons are replaced by spin-selective pulses that
 target \Htwo. Such pulses may be readily obtained for glutamine and Glu at
high field ($\sim 10$T), where the chemical shift between \Htwo\ and the nearby
 proton (\Hfour) of about $1.5~\ppm$~\cite{FL08} is sufficient. At lower fields typically used in vivo, selectivity may be obtained by setting the \isotope{H}{1} transmitter frequency downfield from \Htwo\ (by about $1~\ppm$). This
 approach allowed selective addressing of the \Cfour\ methylene protons of $\gamma$-aminobutyric acid (GABA)~\cite{SYCLC04} at 3T and 7T; in that case,
 the chemical shift from the adjacent \Cthree\ proton was only $\sim 1~\ppm$.
 The proton-selective POTENT variant is essential for amino acids, which have an
 additional methine (CH), such as isoleucine, leucine and valine, which have
similar chemical shift between \Htwoa\ and the other protons ($\sim
 1.5$--$2~\ppm$)~\cite{FL08}. Proton signals of all three aliphatic amino acids
were observed in a cerebral abscess by ex vivo proton MRS~\cite{MSLRR10}.

%
%
The original POTENT sequence may be used to 
cool all four labeled carbons of [1,2,3,4-\multisotope{C}{13}{4}]Glu, obtained
in vivo by metabolism of fully labeled glucose~\cite{XS06}. The four labeled
carbons would be cooled to a similar extent by setting the heteronuclear
 refocusing delay to about $\frac{1}{3J}$ (obtained numerically based on data
from ref~\cite{FL08}), such that the efficiency of PT to \Ctwo, \Cthree, and
 \Cfour\ becomes similar (ideally about 90\%). In this case,
 the reset step after the HCC relay would apply to all protons due to the 
similar \Tone\ relaxation times.

%
%
AC may provide spectral editing that resolves the sub-spectrum of a
particular spin from overlapping sub-spectra that originate from other spins.
While PT from a proton spin can enhance the resolution of the target
\isotope{C}{13} spin by at most 4-fold, AC could ideally, in the future, attain an
 additional ten-fold enhancement using 6--7 spins. Such improvement could
 significantly aid in identifying and
quantitating key metabolites that constitute a small percentage of the sample,
which would remain masked, after PT, by overlapping signals of other
 metabolites. 
A powerful strategy for in vivo spectral editing~\cite{SYCLC04} employed
selective addressing and the unique spin-spin coupling of the target spin to
 remove overlapping signals of several metabolites. 
For $\gamma$-aminobutyric acid (GABA), a chemical shift selective saturation 
(CHESS) pulse sequence commonly applied (since the mid-1980s) for water
 suppression\footnote{CHESS applies, in vivo, a selective
 $90^{\circ}$ pulse on the
 target spin (commonly the large water peak at $4.7~\ppm$), followed by three
 simultaneous dephasing gradients (along each axis), pulsed gradients that
 spatially vary the phase of transverse magnetization. CHESS was implemented in
Ref~\cite{SYCLC04} by a hyperbolic secant of narrow bandwidth ($\sim 1.3~\kHz$)
 centered at 3.9 ppm.}~\cite{MSLRR10} suppressed the \Cfour\ methylene and its overlapping
 \isotope{H}{1} signals from creatine, glutathione and macromolecules (all around
 $3~\ppm$); the target GABA \Hfour\ signal was then selectively regenerated by a
novel homonuclear PT from the neighboring \Cthree\ proton (at $1.91~\ppm$) which
 was unaffected by CHESS ($J_{HH} \sim 7~\Hz$). 
AC may provide similar spectral editing for in vivo 13C-MRS by cooling a
 \isotope{C}{13} spin bound to the target spin and transferring the enhanced 
polarization via PT after suppressing the target carbon and all overlapping
 signals.

%
%
The apparent advantage of AC for spectral editing could be diminished when 
considering the limited SNR enhancement achieved by AC with respect to 
multiscan-PT (see section~\ref{sec:AC-sensitivity}).
Consider the scheme whereby the signal of the target carbon is suppressed 
and then regenerated from the nearby carbon enhanced by AC\@. Assuming 
that a single (proton) reset spin was used, AC is limited with respect to 
multiscan-PT (from a proton) over the same duration. Yet, AC might still be 
preferred in practice, due to the lower SAR (see section~\ref{sec:SAR-reduce}). When this scheme is not possible, due to similar coupling patterns 
for the target spin and overlapping spins (of other metabolites), AC may 
selectively enhance the target signal beyond PT\@. However, in this case 
multiscan-PT grants additional selectivity over the other (carbon) 
computation spins according to the \Tone\ ratio, due to the rapid relaxation 
between scans. The overall selectivity of multiscan-PT is therefore similar to 
AC, which is limited by the relaxation time ratio~\cite{EMW11}. 

\subsection{Discussion of the prospects of AC for MRS}
%
Ideally, AC was shown to reach an SNR improvement that is exponential in
the number of spins. However, such cooling is difficult to obtain
 in practice due to several physical limitations, most 
notably (for liquid-state NMR) the limited ratios between the relaxation rate
of the cooled spins and the repolarization rate of the reset 
spins~\footnote{Taking these factors into account, AC could still yield 
significant cooling, particularly in solid-state NMR.}.
For the case of a single reset spin, a common signal-averaging process,
 called here multiscan-PT, can improve the SNR of the target spin more than AC
 over the same period of time. For several reset spins with higher equilibrium
polarization, such as protons, one cycle of AC can cool a computation spin,
such as a \isotope{C}{13} beyond the reset spin polarization. 
 However, the same cooling can in principle be obtained by a more general PT,
 as shown by S{\o}rensen for I$_n$S~\cite{Sorensen89} and 
I$_n$S$_m$~\cite{Sorensen91} spin systems.
In practice, as noted by S{\o}rensen, experimental implementation of such PT
 may not be feasible, in particular when the protons are bound to different
 carbons; in that case, the two steps of PAC (PT and 3B-Comp) may be the only
 intuitive option\footnote{A non-intuitive solution could be obtained by
 numerical optimization}.
 Beyond SNR improvement, AC could contribute to in vivo 13C-MRS in two
 important areas: spectral editing to resolve complex spectra
by selecting for signals of a particular metabolite, and  
reduction of SAR associated with proton decoupling (by reducing the amount
 of data acquisitions).

%
%
Further research is needed to evaluate both potential applications. For
 spectral editing, it is necessary to seek practical scenarios where AC and
HBC are advantageous. HBC can provide direct spectral indication for the
presence of specific (e.g., 1,2-double-labeled) metabolites, while AC generally
 enhances a single spin considerably, ideally much beyond PT\@.
 AC is limited with respect to multiscan-PT, yet it might still be
 advantageous due to the expected reduction in SAR\@. For example, AC could
 enhance spectral editing where the target signal (along with all overlapping
 signals) is suppressed; the target signal could be regenerated by PT from an
adjacent carbon cooled by AC, rather than by PT from a proton. For SAR
 reduction, the expected effect in vivo can be estimated for multiple scans
 using well-established numerical simulations.

%
%
We hope that our results would encourage further research into experimental
 open system cooling in liquid NMR\@. It appears that there is real benefit for 
\emph{in vivo} heteronuclear brain spectroscopy, a noninvasive method that provides
 unique information about metabolism in specific compartments of the brain.
To this end, further evaluation is needed for the prospective near-future
 applications identified in this work -- spectral editing and SAR reduction.
%
%

\printbibliography

\end{document}